\documentclass[twocolumn]{aastex63}

\usepackage{graphicx}
\usepackage{color, colortbl}
\usepackage{float}

\definecolor{Red}{rgb}{1,0,0}
\definecolor{Blue}{rgb}{0,0,1}
\definecolor{Green}{rgb}{0,1,0}
\definecolor{Purple}{rgb}{1,0,1}

\submitjournal{\apj}

\shorttitle{XTE J1908$+$094}
\shortauthors{Draghis et al.}

\begin{document}

\title{The Spin and Orientation of the Black Hole in XTE J1908$+$094}

\author[0000-0002-2218-2306]{Paul A. Draghis}
\email{pdraghis@umich.edu}
\affiliation{Department of Astronomy, University of Michigan, 1085 South University Avenue, Ann Arbor, MI 48109, USA}

\author{Jon M. Miller}
\affiliation{Department of Astronomy, University of Michigan, 1085 South University Avenue, Ann Arbor, MI 48109, USA}

\author{Abderahmen Zoghbi}
\affiliation{Department of Astronomy, University of Michigan, 1085 South University Avenue, Ann Arbor, MI 48109, USA}

\author{Elias S. Kammoun}
\affiliation{Department of Astronomy, University of Michigan, 1085 South University Avenue, Ann Arbor, MI 48109, USA}
\affiliation{IRAP, Universite de Toulouse, CNRS, UPS, CNES 9, Avenue du Colonel Roche, BP 44346, F-31028, Toulouse Cedex 4, France}

\author{Mark T. Reynolds}
\affiliation{Department of Astronomy, University of Michigan, 1085 South University Avenue, Ann Arbor, MI 48109, USA}

\author{John A. Tomsick}
\affiliation{Space Sciences Laboratory, 7 Gauss Way, University of California, Berkeley, CA 94720-7450, USA}

\begin{abstract}
\textit{NuSTAR} observed the black hole candidate XTE J1908$+$094 during its 2013 and 2019 outbursts. We use relativistic reflection to measure the spin of the black hole through 19 different assumptions of \texttt{relxill} flavors and parameter combinations. The most favored model in terms of Deviance Information Criterion (DIC) measures the spin of the black hole to be $a = 0.55^{+0.29}_{-0.45}$, and an inclination of $\theta=27^{+2}_{-3}$ degrees ($1\sigma$ statistical errors). We look at the effects of coronal geometry assumptions and density of the accretion disk on the spin prediction. All 19 tested models provide consistent spin estimates. We discuss the evolution of spin measurement techniques using relativistic reflection in X-ray binaries and discuss the implications of this spin measurement in reconciling the distributions of stellar mass black hole spin measurements made through X-ray and gravitational wave observations.
\end{abstract}


\section{Introduction} \label{sec:intro}

%% 2-3 paragraphs about relativistic reflection
The two main methods for measuring black hole spin in X-ray binaries are disk continuum fitting (see e.g., \citealt{2006ApJ...652..518M, 2014ApJ...784L..18M, 2014ApJ...793L..29S}) and relativistic reflection (e.g., \citealt{2007ARA&A..45..441M, 2015PhR...548....1M, 2020arXiv201108948R}). Spin measurements made using continuum fitting are strongly dependent on prior knowledge about the mass of the black hole, the distance to the system, the inclination of the inner accretion disk, and the mass accretion rate due to the degeneracy between these parameters in the model. In contrast, the relativistic reflection method requires no prior knowledge about the mass, distance, and mass accretion rate of the black hole in the system, with the inclination of the inner disk being treated as a free parameter. Currently, the main assumption of relativistic reflection models is the nature of the compact corona in the system, with some models assuming a ``lamp-post'' coronal geometry, while others make no prior coronal geometry assumptions. Therefore, relativistic reflection can be applied in systems where, e.g., foreground extinction may inhibit binary mass constraints and is applicable in a larger number of sources.

The two main features of relativistic reflection are the Fe K fluorescence line and a broad flux excess between 20--40~keV known as the ``Compton hump''. The Fe K spectral line is present at 6.4~keV for neutral gas and at progressively higher energies for ionized gas, up to 6.97~keV for H-like Fe XXVI. The line profile of the Fe K line originating from matter in the proximity of a black hole is ``blurred'' by relativistic Doppler shifts and gravitational red-shifts. This method works under the assumption of an optically thick, geometrically thin accretion disk (\citealt{1973A&A....24..337S}) that extends all the way to the Innermost Stable Circular Orbit (ISCO). Numerical simulations show that for an Eddington fraction $\leq0.3$, the assumption of an accretion disk extending to the ISCO holds (\citealt{2008ApJ...676..549S}) and that any gas at smaller radii is infalling onto the black hole and is optically thin (\citealt{Reynolds_2008}). As the size of the ISCO is dependent on the black hole spin in the Kerr metric (\citealt{1972ApJ...178..347B, 1973blho.conf..343N}), measuring the blurring of the Fe K line allows an independent measure of the black hole spin. 

%% paragraph about linking with LIGO results
In the emerging era of gravitational wave (GW) signals from binary black hole (BBH) mergers, the issue of breaking the degeneracy between the mass ratio $q=M_2/M_1$ in the binary and the effective spin parameter $\chi_{\rm eff}=\frac{M_1\chi_1+M_2\chi_2}{M_1+M_2}$  (where $M_{1,2}$ are the masses of the components in the binary and $\chi_{1,2}$ are the components of the spin aligned with the orbital angular momentum) is prohibiting precise measurements of black hole spin (see, e.g., \citealt{2016PhRvD..93h4042P, 2018ApJ...868..140T}). An additional degeneracy between the spins of the two black holes is further complicating the spin measurement (\citealt{2016PhRvD..93h4042P}). As the posterior distributions of GW spin measurements are correlated to the assumed prior distribution (\citealt{2017PhRvL.119y1103V, 2019PhRvX...9c1040A, 2020arXiv201014527A}), it is important to have educated predictions for the prior distribution of spins (\citealt{2020ApJ...899L..17Z}). The most pragmatic choice for an informative prior distribution comes from a robust measurement of the spins of stellar-mass black holes in X-ray binary systems. 

%% 2-3 paragraphs about XTE J1908
The black hole system XTE J1908$+$094 was discovered in February 2002 using the \textit{Rossi X-ray Timing Explorer (RXTE)} Proportional Counter Array (PCA) (\citealt{2002IAUC.7856....1W}). Early spectra indicated the presence of broadened Fe line emission, and \citealt{2009ApJ...697..900M} measured a spin of $a=0.75\pm0.09$ using \textit{BeppoSAX} MECS spectra. Since its discovery, the source has undergone three outbursts, in 2003, 2013, and 2019. The 2013 and 2019 outbursts have been observed with \textit{NuSTAR} (\citealt{2013ApJ...770..103H}). During the 2013 outburst, \citealt{2017MNRAS.468.2788R} have detected resolved, expanding radio jets originating from XTE J1908$+$094 using VLBA and EVN. 

\citealt{2015ApJ...811...51T} and \citealt{2015ApJ...813...90Z} have indicated the presence of a broadened Fe K line in the 2013 \textit{NuSTAR} spectra of XTE J1908$+$094. Still, both papers concluded that spin measurements cannot be made using the dataset. The 2019 \textit{NuSTAR} observation was analyzed in \citealt{2021arXiv210406453C}, but no attempt at a spin measurement was made. Motivated by the characteristics of \textit{NuSTAR} such as its wide pass band and its high sensitivity and by previous similar spin measurements using relativistic reflection (see e.g., \citealt{King_2014, 2016ApJ...826L..12E, 2020ApJ...900...78D, 2020ApJ...893...30X}), we revisited the 2013 and 2019 observations of XTE J1908$+$094. In Section \ref{sec:obs} we describe our data reduction process, in Section \ref{sec:analysis} we describe our analysis and results, and in Section \ref{sec:disc} we discuss the implications of the result, connecting it to the future of spin measurements in X-ray binaries and BBH.

\newpage

\section{Observations and Data Reduction} \label{sec:obs}
\begin{figure}[h!]
    \centering
    \includegraphics[width=0.49\textwidth]{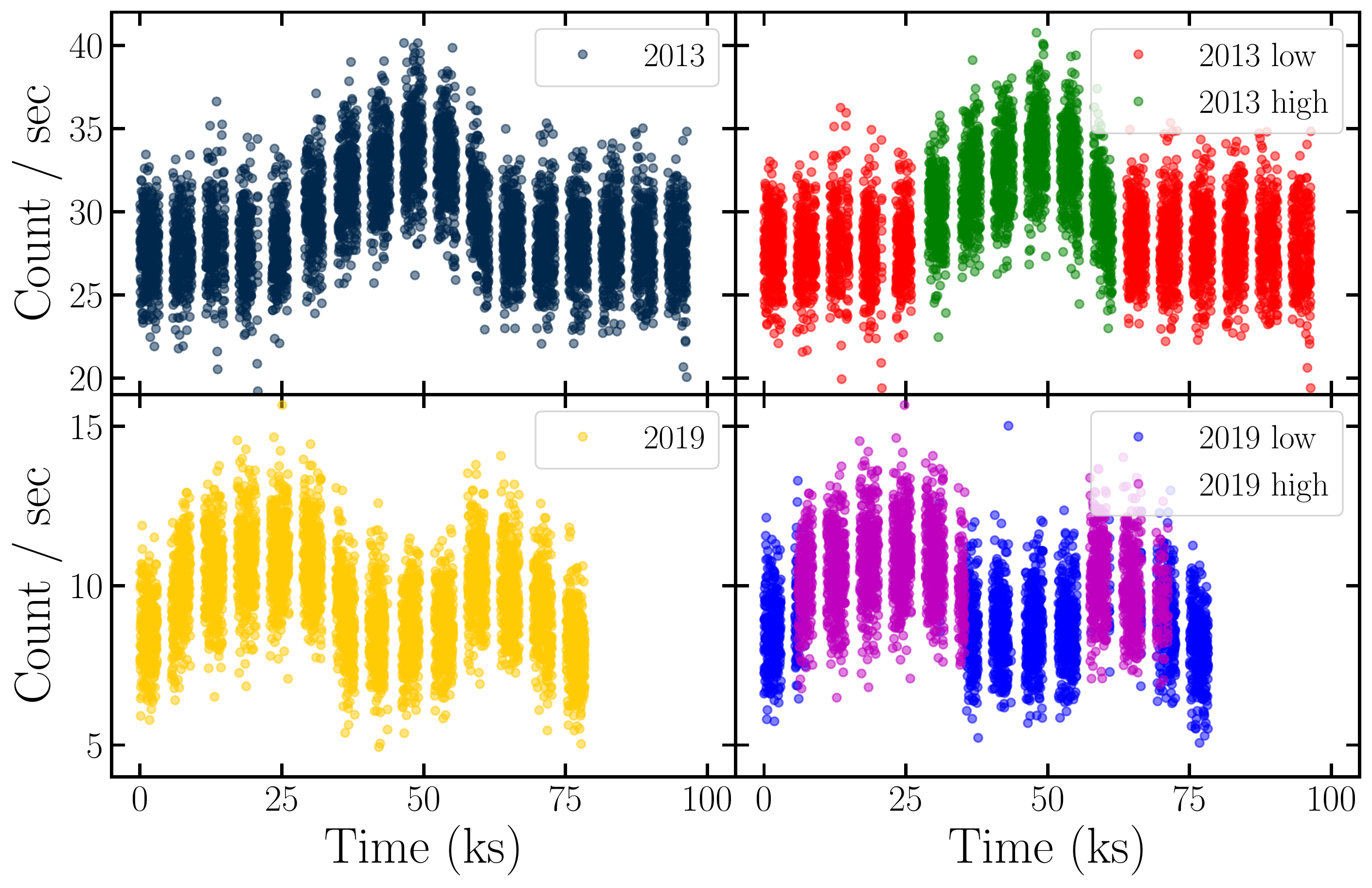}
    \caption{Light curves of the two analyzed observations in the 3--79~keV energy band. The light curves are binned so that each point represents 10 seconds. The left panels show the light curve of the entire observation, while the right panels show the light curves split according to their hardness, as they were considered in the extraction of the spectra that were used for the rest of the analysis. The top panels correspond to the 2013 observation, while the bottom panels represent the 2019 observation.}
    \label{fig:light_curve}
\end{figure}

\textit{NuSTAR} observed XTE J1908$+$094 on 2013 November 8 under ObsID 80001014002 for a net exposure of 45 ks and on 2019 April 10 under ObsID 90501317002 for 41 ks.  During both observations, the source was in the soft state (\citealt{2013ATel.5549....1N, 2019ATel12652....1L}). The data were reduced using the routines in HEASOFT v6.27.1 through the NuSTARDAS pipeline v1.9.2 and CALDB v20200510. The source spectra were extracted from circular regions centered on the source position with radii of 100" in the two FPM NuSTAR sensors, and regions of the same size were used for extraction of background rates. The spectra were grouped using the ``ftgrouppha'' ftool, through the optimal binning scheme described by \citealt{2016A&A...587A.151K}. One potential issue with this binning scheme would be a small number of counts in some bins, particularly towards high energies, for which $\chi^2$ statistics would not be appropriate. Therefore, we compared the fits of our models on the spectra binned with the optimal scheme with the same spectra when grouped by oversampling the resolution by a factor of 3 and requiring a minimal signal to noise ratio of 6. The best fit parameters were strongly consistent between binning schemes. The left panels in Figure \ref{fig:light_curve} show the light curve of the two NuSTAR observations. For the 2013 observation (top left panel), the central part of the light curve shows a 20\% increase in count rate over the 28 c/s continuum, while the 2019 observation (bottom left panel) shows variability on the order of 40\%. 

\begin{figure}[h!]
    \centering
    \includegraphics[width=0.49\textwidth]{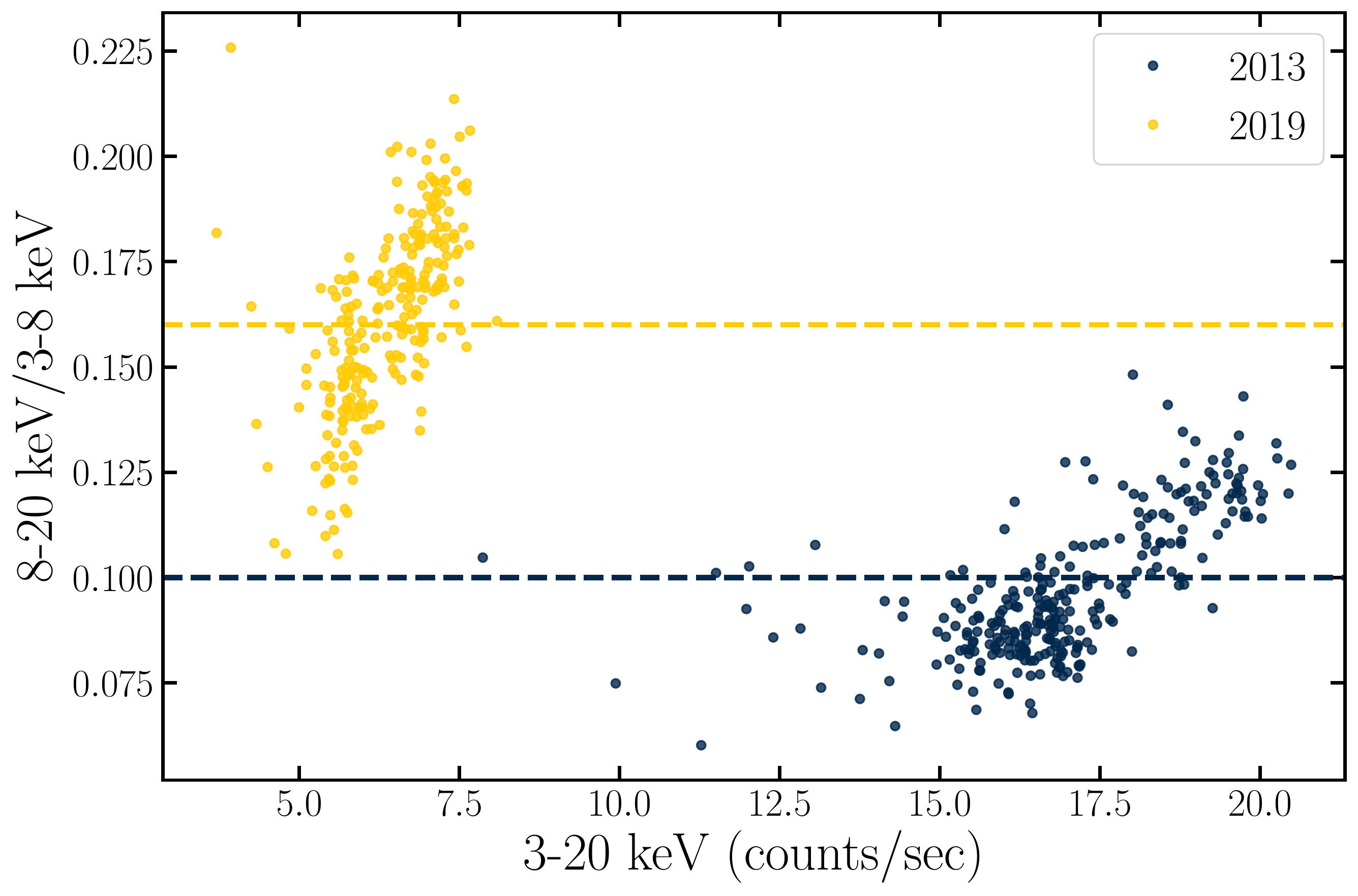}
    \caption{The hardness ratio throughout the two observations computed as the ratio of the 8--20~keV count rate to the 3--8~keV count rate, as a function of the total 3--20~keV count rate. The blue points represent the 2013 observation, while the yellow points represent the 2019 one. The horizontal dashed lines show the hardness cuts placed on the two observations to divide them into the ``high hardness'' and ``low hardness'' spectra. The light curves were binned so that each point represents 200 seconds.}
    \label{fig:hardness}
\end{figure}

For both observations, we extracted the light curves in the 3--8~keV and 8--20~keV bands and compared the ratio of the count rates in the two energy bands to the total count rate in the 3--20~keV band. The lower limit of 3~keV was chosen as the lower bound of the \textit{NuSTAR} pass band. The upper limit of 20~keV was chosen to be just under the Compton hump. The 8~keV break point was chosen as to separate the effects of the Fe complex and the disk emission from coronal emission typical of X-ray binary systems. This hardness ratio is shown in Figure \ref{fig:hardness}. For both observations, the instantaneous hardness appears to have a bimodal distribution. Therefore, we split the two observations into regions with hardness greater and lower than 0.16 for the 2019 observation and greater/lower than 0.1 for the 2013 observation. We generated GTI files using the XSELECT tool by filtering the intensity as described above, and re-extracted the spectra of the low and high hardness intervals of each observation. The right panels in Figure \ref{fig:light_curve} show the intervals of the light curve corresponding to the high and low hardness part of the observations. For the 2013 observation, the high hardness interval corresponds to the central increase in overall flux, and results in the same division of the spectrum as that used by \citealt{2015ApJ...811...51T, 2015ApJ...813...90Z}. Similarly, for the 2019 observation the increase in hardness is associated with an increase in total count rate of the source. The produced spectra are shown in panel (a) of Figure \ref{fig:delchi}. We used the entire NuSTAR energy band (3--79~keV) throughout the spectral fits. 

Upon the completion of the analysis, we tested the robustness of the split of the observations into hardness regimes by re-extracting the spectra after redefining the hardness ratio as the count rates in the 3--6~keV and 6--20~keV bands. We note that the final results are strongly consistent with the initial measurements, implying that in this case, the choice of definition of hardness ratio has little to no impact on the spin measurement.

\section{Analysis and Results} \label{sec:analysis}
The spectral fitting was carried out using XSPEC v12.11.0m (\citealt{1996ASPC..101...17A}), with the quality of the fit being evaluated using $\chi^2$ statistics. The CALDB version used accounts for the low energy excess of the spectra produced by the NuSTAR FPMA sensor due to the tear in its thermal blanket, described by \citealt{2020arXiv200500569M}. While a multiplicative constant is often allowed to vary between the spectra from the two FPM \textit{NuSTAR} sensors, in this particular case the data did not require it. By splitting the two observations into two hardness regimes, we obtain four data groups, each consisting of the FPMA/B spectra, fit jointly.

Visually inspecting the spectra in panel (a) of Figure \ref{fig:delchi} suggests that, on the zeroth order, models need at least two components to accommodate for the apparent slope change at $\sim9$~keV. We fit the spectra with the combination of a disk blackbody component \texttt{diskbb} (\citealt{1984PASJ...36..741M}), characteristic of accretion disks around compact objects, and a power law.  To account for interstellar absorption, we included the multiplicative component \texttt{TBabs} (\citealt{2006HEAD....9.1360W}), with the abundances computed according to \citealt{2000ApJ...542..914W} and photoionization cross-sections computed by \citealt{1996ApJ...465..487V}. The residuals of this model are shown in panel (b) of Figure \ref{fig:delchi}, with the fit producing $\chi^2/\nu=2740.25/1621=1.69$.  The main unmodeled feature is a strong emission line consistent with the Fe K line complex. The decrease in flux both below and above the Fe K line suggests that the fitting procedure altered the continuum shape of the disk blackbody in order to balance the negative and positive residuals, therefore impacting the shape of the continuum at low energy and inducing an apparent flux excess in the 3-4~keV range. Additionally, a broad excess is seen in the 20--40~keV range. These features are consistent with relativistic disk reflection.

For the measured disk temperature is $kT_{\rm in}\sim$0.7~keV, the energy distribution peaks around $\sim$ 2~keV, just below the lower limit of the \textit{NuSTAR} pass band.  As the high energy tail of the disk continuum easily extends into the \textit{NuSTAR} band, the ability to constrain interstellar absorption and the temperature of the accretion disk would not be improved significantly through the addition of simultaneous observations using the \textit{Swift} X-ray telescope (XRT; \citealt{2007SPIE.6686E..07B}). \citealt{2015ApJ...813...90Z} pointed out that when fitting the \textit{Swift} and \textit{NuSTAR} spectra from the same day simultaneously, the residuals show a difference between the spectra from the two instruments in the overlapping energy band. As \textit{Swift} detectors suffer from pile-up, the reduced sensitivity obtained when extracting annuli does not permit improving the constraints placed on parameters.

\subsection{The \texttt{relxill} family of models}
\begin{figure}[h!]
    \centering
    \includegraphics[width=0.49\textwidth]{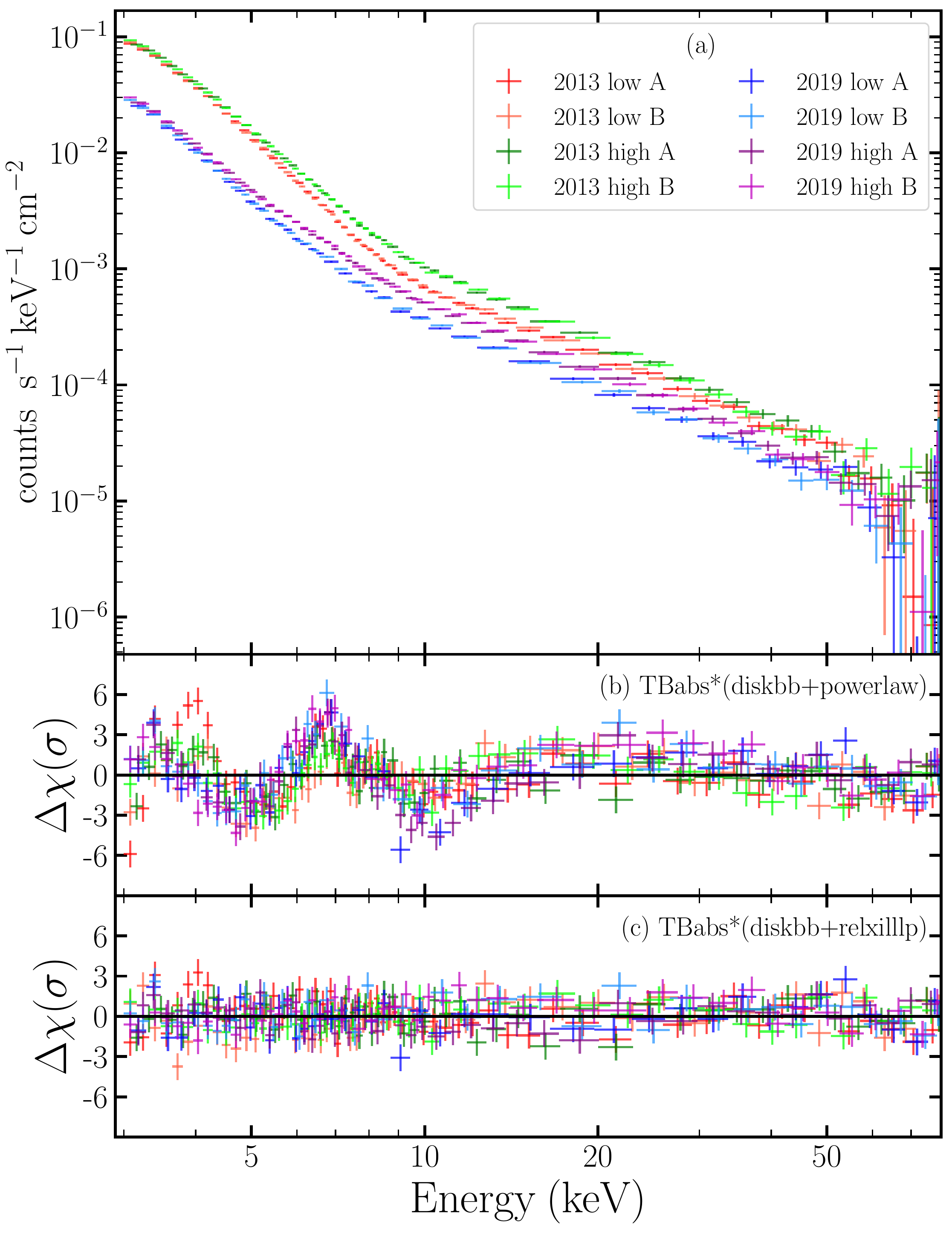}
    \caption{Panel (a) shows the spectrum of XTE J1908$+$094. The red, green, blue, and purple points represent the 2013 low hardness, 2013 high hardness, 2019 low hardness and 2019 high hardness data groups respectively. Panel (b) shows the residuals in terms of $\sigma$ for the \texttt{TBabs*(diskbb+powerlaw)} model. Panel (c) shows the residuals of the \texttt{TBabs*(diskbb+relxilllp)} model, which is the best performing model of the 19 tested in terms of DIC (model 6).}
    \label{fig:delchi}
\end{figure}

The relativistic reflection features were modeled by replacing the \texttt{powerlaw} component of the model with multiple models that are part of the \texttt{relxill} family of models (\citealt{2014MNRAS.444L.100D, 2014ApJ...782...76G}). We used \texttt{relxill} v1.4.0. Panel (c) of Figure \ref{fig:delchi} shows the residuals of one of such models, with the reflection component being modeled by \texttt{relxilllp} (described below), thus making the full model \texttt{TBabs*(diskbb+relxilllp)}.  The quality of the fit is drastically improved, producing $\chi^2/\nu=1707.44/1604=1.06$. 

The \texttt{relxill} family of models is now a standard tool in modeling relativistic reflection. Variants of \texttt{relxill} allow probing the effects on the spectrum of different physical assumptions, such as the geometry of the corona, density and ionization of the accretion disk, iron abundance, geometry of the accretion disk, and spin of the black hole. As the goal of this analysis is to measure the spin of the black hole, $a = cJ/GM^{2}$ ($-1 \leq a \leq 1$), we aimed to explore the entire range of parameter combinations and theoretical assumptions that the \texttt{relxill} family of models provides.

All \texttt{relxill} models include the spin of the black hole as a free parameter. The geometry of the accretion disk is constrained through the inner and outer radius of the disk ($r_{\rm in}$ and $r_{\rm out}$) and the inclination of the inner accretion region of the accretion disk. For our analysis, we fixed the inner radius of the accretion disk at the ISCO of the black hole, as this assumption is in agreement with MHD simulations such as those by \citealt{Reynolds_2008}. The outer radius of the disk was fixed at a large value of $r_{\rm out}=990\;r_g$. The inclination of the inner accretion disk was allowed to vary as a free parameter. Note that the inclination of the inner disk does not necessarily match that of the outer regions of the accretion disk and of the binary system, as a result of the Bardeen-Petterson effect (see, e.g., \citealt{1975ApJ...195L..65B, 2015MNRAS.448.1526N, 2020MNRAS.tmp..707L}). Still, the orientation of the black hole spin and the inner accretion disk can precess due to the Lense-Thirring effect and align itself with the outer accretion disk (see, e.g., \citealt{2019MNRAS.487.3488B}).

The iron abundance of the reflector measured relative to the solar abundance $A_{\rm Fe}$ and the ionization of the reflector ($\xi=L/n r^2$, with $L$ being the luminosity of the emitting source, $n$ the hydrogen number density of the reflector, and $r$ the distance between source and reflector) are allowed to vary in all models. The illuminating flux is modeled as a power law with index $\Gamma$ and a high energy cutoff $E_{\rm cut}$. The power law index is a free parameter in all models, while the high energy cutoff is free in models having a fixed density of the accretion disk of $n=10^{15}\;cm^{-3}$ (\texttt{relxilllp} and \texttt{relxill}), and fixed by construction of the models to $E_{\rm cut}=300$~keV in models that allow higher densities of the accretion disk (\texttt{relxillD}).

An alternative description of the emissivity of the corona is to model it as a thermal Comptonization continuum. The \texttt{relxill} family of models does offer this possibility through the \texttt{relxillCp} and \texttt{relxilllpCp} models. In order for black hole coronae to produce the observed X-ray fluxes while being powered by hot electrons, the size of coronae needs to be on the order of $\sim 1000~\rm r_g$ (\citealt{2001MNRAS.321..549M}). A corona of this physical size cannot produce short term variability on the timescales observed in some sources (see e.g., \citealt{2000MNRAS.318..857L, 2015MNRAS.451.4375F}). \citealt{2001MNRAS.321..549M} suggested that the energy must be stored in the corona as magnetic fields generated by a sheared rotator. Another possibility is a combination of the two: the compact base of a jet surrounded by a larger cloud of hot electrons. As a test, we compare our best model (see below, Subsection \ref{subsec:relxilllp}) with the equivalent Comptonization model. The fit becomes worse by $\Delta \chi^2\approx10$ for no change in the number of degrees of freedom, while the best fit parameters remain strongly consistent with the initial results. As the shape of the emissivity of the corona does not appear to influence the spin measurement, we continued to only analyze models that use a power law emissivity.

In the \texttt{relxilllp} model, the geometry of the corona is modeled as a compact emitter at height $h$ above the accretion disk (the ``lamp post'' assumption). In \texttt{relxill} and \texttt{relxillD}), the geometry of the corona is unconstrained, but its emissivity is modeled as a broken power law with radius ($J\propto r^{-q_1} \text{ for } r<r_{\rm break}$, and $J\propto r^{-q_2} \text{ for } r>r_{\rm break}$). Additionally, all \texttt{relxill} models have a parameter representing the normalization of the flux of the component ($\rm norm_r$) and a ``reflection fraction'' parameter ($\rm refl\_frac$) which represents the ratio of the intensity emitted towards the accretion disk and the intensity escaping to infinity in the frame of the primary source (\citealt{2016A&A...590A..76D}). For models assuming the lamp post geometry, this is calculated using ray tracing simulations as described by \citealt{2014MNRAS.444L.100D}. In this version of the model, for the \texttt{relxill} variants that do not assume a coronal geometry, the reflection fraction is approximated by taking the ratio of the reflected flux to the direct flux in the 20--40~keV band (\citealt{2016AN....337..362D}).

The \texttt{relxilllpion} variant of \texttt{relxill} replaces the constant ionization parameter with a ionization gradient modeled as a power law with radius. With this model, the solution converges to the same parameter combination as in the case with constant ionization, with the power law index for the ionization gradient being consistent with zero. While this does not exclude the possibility of an ionization gradient across the accretion disk, this data set is not able to constrain any possible variation in ionization. Therefore, we chose not to explore models using variable ionization further. The lack of an apparent ionization gradient is consistent with a steep emissivity profile. A flatter emissivity would include more emission from large radii which could, in principle, have lower ionization. This would make the effects of a ionization gradient more apparent.

\subsection{MCMC analysis}
We explored a total of 19 variations of the above models. For each individual combination of model and parameter assumptions, we ran the default XSPEC fitting algorithm to find the ``best fit'' parameter combination that minimizes the $\chi^2$ statistic given the data. This set of parameters was used to construct a proposal distribution for running a Monte Carlo Markov Chain (MCMC). The proposal distribution was generated from a Gaussian distribution around the ``best fit'' parameters. The chains were run with 200 walkers for a total of $2\times10^6$ steps, using the XSPEC EMCEE implementation written by A. Zoghbi\footnote{ \url{https://zoghbi-a.github.io/xspec\_emcee/} }. The number of walkers was chosen as an integer on the order of a few times the number of free parameters. \citealt{Sokal1996MonteCM} suggested running MCMC chains with more steps than $\sim 1000\times \tau_f$, where $\tau_f$ represents the integrated autocorrelation time. We computed $\tau_f$ using the prescription given by \citealt{2010CAMCS...5...65G} and obtained that running the chains for $2\times10^6$ steps is equivalent to $3-4000\times\tau_f$, ensuring convergence of the chains. The first $5\times10^5$ steps of the chains were then assumed to be a ``burn in'' phase and therefore disregarded.

Using the MCMC posterior distribution, we computed the Deviance Information Criterion (DIC, \citealt{DIC_text}) for each model. We used the DIC to quantify the goodness of the fit and to distinguish between the tested models. We used this approach instead of an F-test in order to properly account for how well the model fits the data and how complex the model is, all while not only looking at one single parameter combination of maximum likelihood, but at the entire distribution around the best fit value. Table \ref{table:models} in Appendix \ref{sec:models} presents the entire set of models tested, together with the \texttt{relxill} variant used, the assumptions made, the best-fit $\chi^2$ and number of degrees of freedom, reduced $\chi^2$, computed DIC, and model ranking based on the DIC. 

\begin{figure}[h!]
    \centering
    \includegraphics[width=0.49\textwidth]{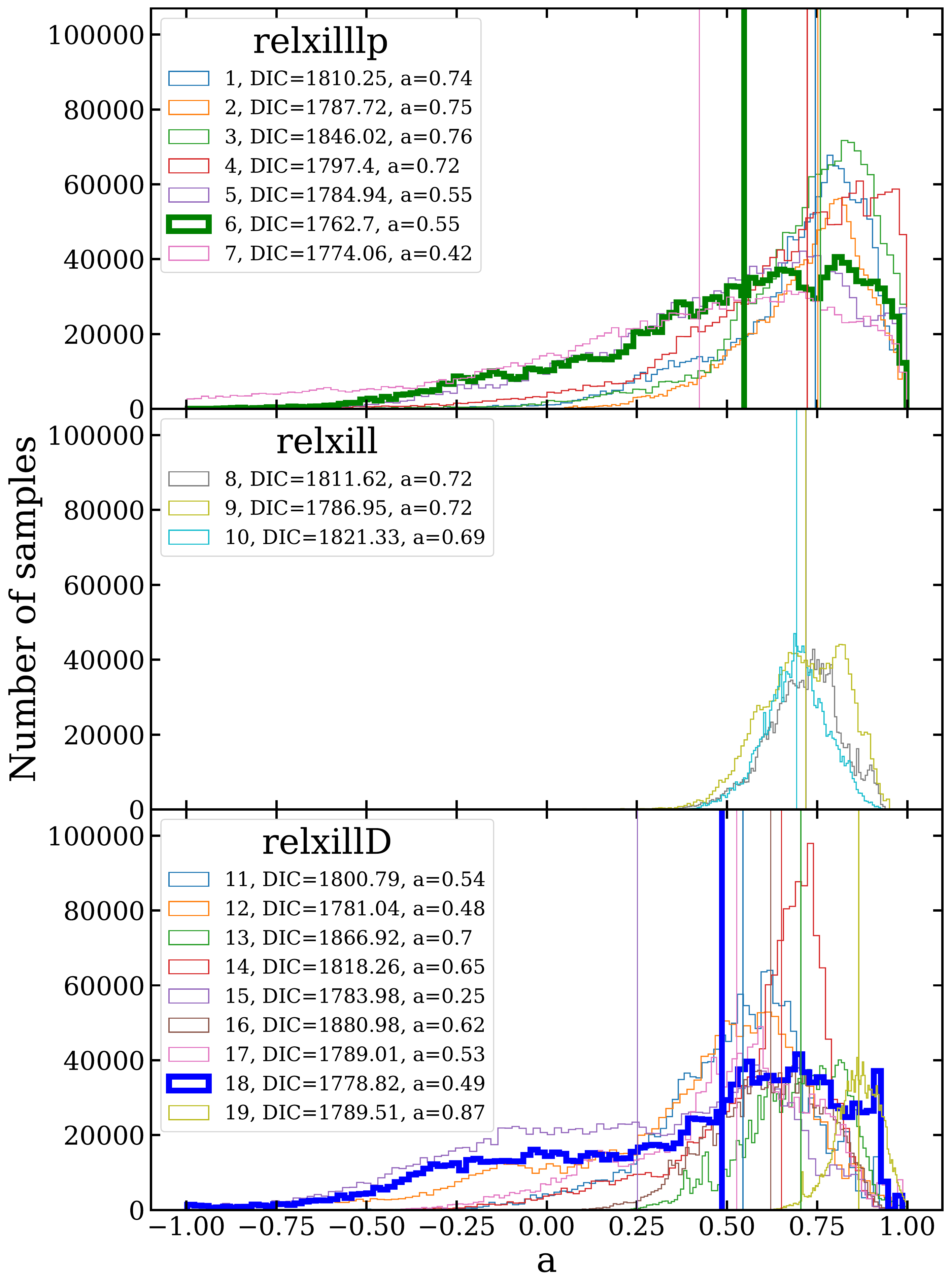}
    \caption{Histogram of the posterior distributions of the black hole spin parameter in all 19 tested models. The top panel shows the models using \texttt{relxilllp}. The middle panel shows the models using \texttt{relxill}, and the bottom panel show models using \texttt{relxillD}. The vertical lines represent the median value of the spin posterior distribution for each model. The highlighted green line in the top panel represents the best performing ``lamp post'' model in terms of DIC (model 6). The blue curve in the bottom panel shows the model performing best in terms of DIC from the \texttt{relxill} variants that adopt no assumptions about the geometry of the corona.}
    \label{fig:histograms}
\end{figure}

Figure \ref{fig:histograms} shows the histograms of the posterior distributions of the MCMC chains for the black hole spin for the 19 tested models. The top panel shows the model variants that use \texttt{relxilllp} to model the reflection component, the middle panel shows the distributions for the variants using \texttt{relxill}, and the lower panel shows the models using \texttt{relxillD}. The models are presented below. The vertical lines represent the median of the posterior distribution for the spin parameter in each model. The highlighted green curve in the top panel and the blue curve in the bottom panel correspond to models 6 and 18, presented in detail below. The spin measurements are consistent between all models, suggesting a robustness of the measurement against the assumed \texttt{relxill} flavor used.

\subsection{Best ``lamp-post'' model}\label{subsec:relxilllp}
Based on the computed DIC, the model that best fits the data in the lamp post assumption is that where the height of the corona is linked between the low and high hardness spectra from the same observation (model 6 in \ref{table:models}). Table \ref{table:results_lp} shows the median values across the posterior sample in the MCMC chains with $1\sigma$ uncertainties. For the black hole spin, the number in parentheses represent the value at which $\chi^2$ is minimized. The residuals of the fit using this model are shown in panel (c) of Figure \ref{fig:delchi}. The number in parentheses in the $\chi^2/\nu$ row represents the median $\chi^2$ value across the posterior distribution from the MCMC chains, with its $1\sigma$ variation. The measured spin of the black hole for the best performing model in terms of DIC is $a=0.55^{+0.29}_{-0.45}$.  The posterior distribution of the spin parameter in the MCMC chain is shown in the top panel in Figure \ref{fig:histograms} by the highlighted green line. This measurement is consistent with all other measurements made using the other tested models. 

Model 6 links the height of the compact corona in the \texttt{relxilllp} within the 2013 and 2019 observations. The measured values for the height of the corona are $h_{\rm 2013}=14^{+2}_{-2}~ r_g$ for the 2013 observation and $h_{\rm 2019}=6^{+1}_{-1}~r_g$ for the 2019 observation. If instead we link the height of the corona between all four data groups, the measured value is the average of the previous two measurements, $h=10^{+1}_{-1}~ r_g$. This corresponds to model 7 in Table \ref{table:models}, which produces the second best DIC of the 19 models. Allowing the height of the corona to be free in all data groups (model 5) produces a similar $\chi^2$ value at the cost of two extra free parameters, resulting in a worse DIC, ranking 6th among the 19 tested models. The measurements for the height of the corona in this model are $h_{\rm 2013\;low}=14^{+3}_{-2} ~ r_g$, $h_{\rm 2013\;high}=15^{+3}_{-2}~ r_g$, $h_{\rm 2019\;low}=6^{+1}_{-1}~ r_g$, and $h_{\rm 2019\;high}=6^{+1}_{-1}~ r_g$, consistent with the values obtained when the height of the corona is linked within each observation. This result suggests that in the lamp post geometry, the corona of XTE J1908$+$094 moves from $\sim14~r_g$ in 2013 to $\sim6~r_g$ in 2019. 

\begin{deluxetable}{c|c|c|c|c}
\tablecaption{Results from Model 6 \label{table:results_lp}}
\tablewidth{\textwidth} 
\tabletypesize{\scriptsize}
\tablehead{
\colhead{Parameter} & 
 \colhead{\cellcolor{Red} 2013 low} & 
 \colhead{\cellcolor{Green} 2013 high} & 
 \colhead{\cellcolor{Blue} 2019 low} & 
 \colhead{\cellcolor{Purple} 2019 high} 
}
\startdata\\
$N_{\rm H} \: (\rm \times 10^{22} \: cm^{-2})$ & \multicolumn{4}{c}{$3.6^{+0.1}_{-0.1}$} \\ \hline
$a$ & \multicolumn{4}{c}{$0.55^{+0.29}_{-0.45}\:(0.71)$} \\
$\rm Incl \: (^\circ)$ & \multicolumn{4}{c}{$27^{+2}_{-3}$} \\
$A_{\rm Fe}$ & \multicolumn{4}{c}{$6^{+2}_{-1}$} \\ \cline{2-5} 
$h \: (\rm r_g)$ & \multicolumn{2}{c|}{$14^{+2}_{-2}$} & \multicolumn{2}{c}{$6^{+1}_{-1}$} \\ \cline{2-5} 
$kT_{\rm in} \: (\rm keV)$ & $0.709^{+0.003}_{-0.003}$ & $0.714^{+0.003}_{-0.004}$ & $0.598^{+0.004}_{-0.004}$ & $0.606^{+0.005}_{-0.005}$ \\
$\rm norm_{\rm d} \: (\rm \times 10^{3})$ & $1.25^{+0.04}_{-0.04}$ & $1.17^{+0.04}_{-0.04}$ & $1.07^{+0.06}_{-0.05}$ & $0.98^{+0.06}_{-0.05}$ \\
$\Gamma$ & $2.08^{+0.02}_{-0.03}$ & $2.21^{+0.02}_{-0.02}$ & $2.12^{+0.02}_{-0.03}$ & $2.16^{+0.02}_{-0.02}$ \\
$\rm Log(\xi)$ & $3.8^{+0.1}_{-0.1}$ & $4.3^{+0.1}_{-0.1}$ & $3.9^{+0.1}_{-0.1}$ & $4.0^{+0.1}_{-0.1}$ \\
$E_{\rm cut}\: (\rm keV)$ & $500^{+200}_{-200}$ & $700^{+200}_{-200}$ & $600^{+200}_{-200}$ & $700^{+200}_{-200}$ \\
$\rm refl \_ frac$ & $0.6^{+0.1}_{-0.1}$ & $0.7^{+0.1}_{-0.1}$ & $0.9^{+0.2}_{-0.2}$ & $0.8^{+0.2}_{-0.2}$ \\
$\rm norm_{\rm r} \: (\rm \times 10^{-4})$ & $13^{+1}_{-1}$ & $21^{+2}_{-2}$ & $11^{+2}_{-2}$ & $17^{+3}_{-3}$ \\ \hline
$\chi^2 / \rm bins$ & $531.42/432$ & $398.82/417$ & $376.17/388$ & $401.03/401$ \\ \hline
$\chi^2 / \nu$ & \multicolumn{4}{c}{$1707.44/1604=1.06\:\:\:(1735^{+8}_{-7})$}
\enddata
\tablecomments{In the row presenting the black hole spin, the number in parentheses represents the value that produced the lowest $\chi^2$ in the analysis. In the $\chi^2/\nu$ row, the number in parentheses represents the median $\chi^2$ value in the MCMC analysis with associated $\pm1\sigma$ variation.}
\end{deluxetable}

%% paragraph about high A_Fe
The measured inclination of the inner region of the accretion disk through this model is $Incl=27^{+2}_{-3}$ degrees. The measured Fe abundance is $A_{\rm Fe}=6^{+2}_{-1}$ of the solar value. \citealt{2018ApJ...855....3T} found that for Cygnus X-1, a source with previously reported high Fe abundance ($A_{\rm Fe}\sim10$), fitting the data with a high density variant of the \texttt{reflionx} model (\citealt{2005MNRAS.358..211R}) produces better fits than using \texttt{relxill} variants, even with a solar Fe abundance. For Cygnus X-1, the value of the disk density measured by \citealt{2018ApJ...855....3T} was $n\sim4\times10^{20}\;cm^{-3}$, much higher than the range of densities that the \texttt{relxill} family of models can probe. We were unable to obtain a fit of comparable quality to that of the other tested models when using the model presented in \citealt{2018ApJ...855....3T}. If instead we fix $A_{\rm Fe}=1$ in our tested models, the fits become significantly worse, with an increase of $\chi^2\geq65$ for one fewer degree of freedom across all models. In all models, including the high density variants of \texttt{relxill}, the preferred Fe abundance is high.

\begin{figure}[h!]
    \centering
    \includegraphics[width=0.49\textwidth]{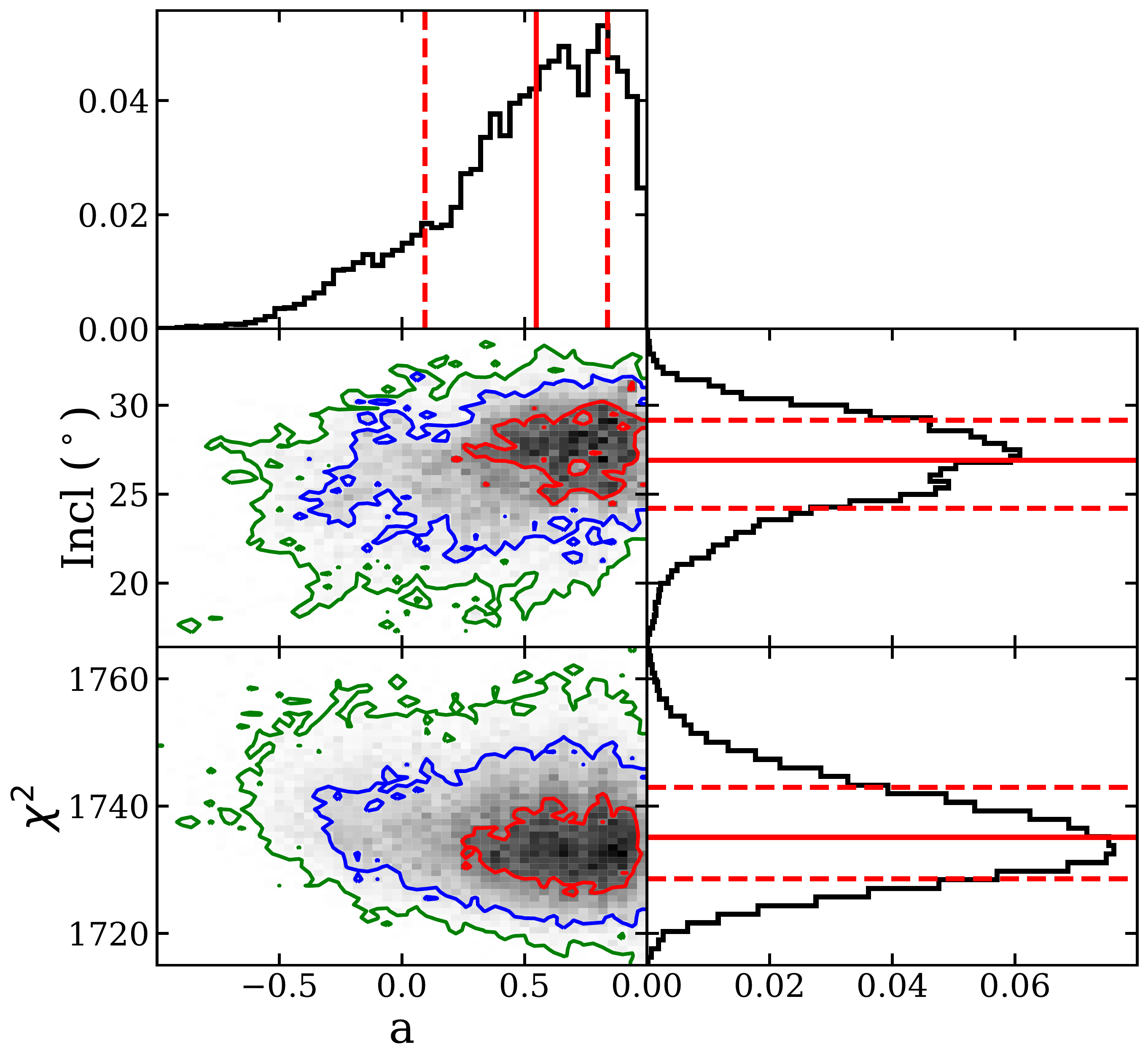}
    \caption{Histogram of the spin-inclination parameter space for model 6 (middle left panel) and of the spin-$\chi^2$ space (bottom left panel). The $1\sigma$, $2\sigma$, and $3\sigma$ confidence intervals are represented by the red, blue, and green contours in the two aforementioned panels. The top left, middle right and bottom right panels show the 1D histograms of the posterior distribution in the MCMC chain of the spin, inclination, and $\chi^2$. The solid red lines represent the median of the distributions and the dashed red lines represent the $\pm1\sigma$ uncertainty around the median.}
    \label{fig:spin_incl_chi}
\end{figure}

As both relativistic Doppler shifts and gravitational redshifts have the effect of producing an asymmetry of spectral lines, the black hole spin parameter and the inclination of the inner accretion disk can be partially degenerate. The middle left panel in Figure \ref{fig:spin_incl_chi} shows the $1\sigma$, $2\sigma$, and $3\sigma$ confidence intervals (in red, blue, and green) of the black hole spin parameter in the posterior MCMC distribution in relation to the inclination of the accretion disk in model 6. The top left and middle right panels show the 1D histograms of the posterior samples in the MCMC run for the spin an inclination of the inner disk. The solid red lines represent the median of the distributions, while the dashed red lines represent the $\pm1\sigma$ confidence limits of the median. We note that no apparent degeneracy is present between the mentioned parameters in this measurement. The bottom left panel in Figure \ref{fig:spin_incl_chi} shows the same confidence intervals for the spin measurement in relation to the $\chi^2$ of the specific parameter combination containing the given spin values. The bottom right panel shows the 1D histogram of the $\chi^2$ values in the MCMC chain. Note that the quality of the fit is not driven mainly by the black hole spin, indicating that error searches across a single parameter can be unreliable. This confirms the need to explore the effects of varying the entire parameter set simultaneously in an error analysis, in order to probe the full set of physical phenomena covered by the model. Besides an apparent degeneracy between the temperature of the \texttt{diskbb} components and the normalization of the component, no other degeneracies are noticeable. For reference, Figure \ref{fig:corner_small} in Appendix \ref{sec:corner} shows the corner plot for a few parameters of interest in the analysis.

%% paragraph about diskbb temp, normalization, and replacing with kerrbb
The emission from the accretion disk was modeled using \texttt{diskbb} (\citealt{1984PASJ...36..741M}). The measured disk temperature in this model is $kT_{\rm in\;2013\;low}=0.709^{+0.003}_{-0.003}$~keV for the low hardness part of the 2013 spectrum, $kT_{\rm in\;2013\;high}=0.714^{+0.003}_{-0.004}$~keV for the high hardness part of the 2013 spectrum, $kT_{\rm in\;2019\;low}=0.598^{+0.004}_{-0.004}$~keV for the low hardness part of the 2019 spectrum, and $kT_{\rm in\;2019\;high}=0.606^{+0.005}_{-0.005}$~keV for the high hardness part of the 2019 spectrum. This is consistent with a disk of temperature $kT_{\rm in\;2013}\sim0.71$~keV for the 2013 observation, $kT_{\rm in\;2019}\sim0.60$~keV for the for the 2019 observation, indicating a constant disk temperature throughout the duration of the observation, but variable over longer periods of time. 

The normalization of the \texttt{diskbb} component is defined as $(R_{in}/D_{10})^2\cos\theta$ where $R_{in}$ is the apparent inner disk radius in km, $D_{10}$ is the distance to the source in units of 10 kpc, and $\theta$ is the inclination angle of the accretion disk. Assuming that the accretion disk extends to the ISCO for both observations, that the distance to the source is constant, and that the inclination does not vary in time, one would expect the normalization of the \texttt{diskbb} component to be constant. Models 1, 2, and 3 fix the normalization of the \texttt{diskbb} component across all data groups under the same assumption about the corona height as models 5, 6, and 7, while maintaining the same reflection component (\texttt{relxilllp)}. Additionally, model 4 fixes the normalization and temperature of \texttt{diskbb} and the height of the corona within one observation. Models 1-4 produce overall worse fits than models 5-7 where the normalization of the accretion disk component is allowed to vary, indicating that the \texttt{diskbb} model does not fully capture the range of physical effects occurring in the accretion disk. If we characterize the emission from the accretion disk using \texttt{kerrbb} (\citealt{2005ApJS..157..335L}) which models an the radiation from a multi-temperature blackbody accretion disk around a Kerr black hole, the fits are similar in quality as when using the \texttt{diskbb} component. Still, the addition of multiple free parameters drives the DIC to much higher values, and the degeneracy between the black hole mass, distance to the source, mass accretion rate and disk inclination cannot be broken without prior knowledge of these parameters. Therefore, we note that while \texttt{diskbb} does not fully describe all physical effects on the emission from accretion disk, its simplicity makes it the preferred method over more complicated models. In practice, the emergent X-ray spectrum from an accretion disk depends on the spectral hardening factor (\citealt{1995ApJ...445..780S}), which must depend on the atmospheric properties of the disk, such as gas density, temperature, ionization state, and magnetization (\citealt{2021MNRAS.500.3640S}). Therefore, it is possible for the disk radius to remain constant, while the inferred radius and disk temperature be observed to vary. 

%% paragraph about ionization and gamma, and how they differ in the low/hard states.
In all \texttt{relxill} models, the illuminating flux is modeled as a power law with index $\Gamma$ and high energy cutoff $E_{\rm cut}$. This model is unable to constrain the high energy cutoff, returning values between $E_{\rm cut}\sim300-900$~keV. The estimated power law index is $\Gamma_{\rm 2013\;low}=2.08^{+0.02}_{-0.03}$ for the low hardness part of the 2013 spectrum, $\Gamma_{\rm 2013\;high}=2.21^{+0.02}_{-0.02}$ for the high hardness part of the 2013 spectrum, $\Gamma_{\rm 2019\;low}=2.12^{+0.02}_{-0.03}$ for the low hardness part of the 2019 spectrum, and $\Gamma_{\rm 2019\;high}=2.16^{+0.02}_{-0.02}$ for the high hardness part of the 2019 spectrum. We note that the power law index $\Gamma$ increases for both observations between the low and high hardness regimes. While this may be contrary to expectations, it is not uncommon, with similar a similar behavior shown by \citealt{2000ApJ...544..993S} in their monitoring of XTE J1550$-$564.

The measured ionization of the accretion disk ($Log\;\xi$) takes high values $\geq3.7$. Similarly to the case of the power law index, the ionization parameter increases during both observations between the low and high hardness parts of the spectrum. Lastly, the measured reflection fraction in \texttt{relxill} is broadly consistent with a value $\sim0.7$ and does not significantly vary between observations.  

\subsection{Best geometry-free model}
%%paragraph about models without assumptions about the corona geometry.
While \texttt{relxilllp} assumes a lamp post coronal geometry, the geometry of the corona is unspecified in \texttt{relxill} and \texttt{relxillD}. Therefore, models 8-19 in Table \ref{table:models} test the impact of the coronal geometry assumption on the spin measurement. Model 8 replaces the \texttt{relxilllp} component with \texttt{relxill}. The parameters describing the emissivity of the corona ($q_1$, $q_2$, and $R_{\rm br}$) are allowed to vary freely. We note that for this model, the breaking radius $R_{\rm br}$ is unconstrained, so model 9 fixes this to a value of $R_{\rm br}=10\;r_g$. Model 10 allows the breaking radius to vary, but links the normalization of the \texttt{diskbb} component due to the arguments mentioned above. \texttt{relxill} allows the high energy cutoff $E_{\rm cut}$ of the power law describing the incident flux to vary, but works under the assumption of an accretion disk density of $Log(n)=15$. To probe the effects of the disk density on the fits, we replaced the \texttt{relxill} component in the model with \texttt{relxillD}. Models 11-13 adopt the same assumptions about coronal emissivity and \texttt{diskbb} normalization as models 8-10 respectively, but with the disk density fixed to $Log(n)=15$. Note that this differs from models 8-10 through the fact that \texttt{relxillD} fixed the high energy cutoff to $E_{\rm cut}=300$~keV, while \texttt{relxill} allows it to vary. Models 14-16 and 17-19 are the analogous of models 11-13 for $Log(n)=17$ and $Log(n)=19$ respectively. 

%%  paragraph about the best performing model
Of the models that make no assumption about the geometry of the corona, model 18 performs best in terms of DIC and ranks third among all 19 models. Model 18 uses the \texttt{relxillD} component, fixes the density of the accretion disk to $Log(n)=19$, and $R_{\rm br}=12~r_g$. We note that, by fixing $R_{\rm br}=12~r_g$, models 18, 12, 15, and 9 perform overall better than the other variants of \texttt{relxill/D} ranking 3rd, 4th, 5th, and 7th of the 19 models in terms of DIC. All these models produce black hole spin measurements that are consistent within the $1\sigma$ uncertainties, as shown by Figure \ref{fig:histograms}. In particular, model 18 predicts a black hole spin of $a=0.5^{+0.3}_{-0.6}$, in good agreement with the result of the best performing lamp post model. The entire set of parameters of model 18 together with their $1\sigma$ uncertainties are shown in Table \ref{table:results_D}, and the histogram of the posterior distribution of the spin parameter in the MCMC chain is shown by the highlighted blue curve in the bottom panel of Figure \ref{fig:histograms}.

\begin{deluxetable}{c|c|c|c|c}
\tablecaption{Results from Model 18 \label{table:results_D}}
\tablewidth{\textwidth} 
\tabletypesize{\scriptsize}
\tablehead{
\colhead{Parameter} & 
 \colhead{\cellcolor{Red} 2013 low} & 
 \colhead{\cellcolor{Green} 2013 high} & 
 \colhead{\cellcolor{Blue} 2019 low} & 
 \colhead{\cellcolor{Purple} 2019 high} 
}
\startdata\\
$N_{\rm H} \: (\rm \times 10^{22} \: cm^{-2})$ & \multicolumn{4}{c}{$3.7^{+0.1}_{-0.1}$} \\ \hline
$a$ & \multicolumn{4}{c}{$0.5^{+0.3}_{-0.6}\:(0.74)$} \\
$\rm Incl \: (^\circ)$ & \multicolumn{4}{c}{$33^{+7}_{-3}$} \\
$A_{\rm Fe}$ & \multicolumn{4}{c}{$6^{+1}_{-1}$} \\
$Log(n)\:(cm^{-3})$ & \multicolumn{4}{c}{$19*$} \\
$R_{\rm br} \: (\rm r_g)$ & \multicolumn{4}{c}{$12*$} \\ \cline{2-5} 
$q_1$ & $6^{+3}_{-3}$ & $6^{+2}_{-2}$ & $5^{+2}_{-2}$ & $6^{+2}_{-2}$ \\
$q_2$ & $2^{+2}_{-1}$ & $3^{+1}_{-1}$ & $3^{+3}_{-2}$ & $3^{+3}_{-2}$ \\
$kT_{\rm in} \: (\rm keV)$ & $0.704^{+0.004}_{-0.004}$ & $0.710^{+0.004}_{-0.004}$ & $0.597^{+0.005}_{-0.005}$ & $0.603^{+0.005}_{-0.005}$ \\
$\rm norm_{\rm d} \: (\rm \times 10^{3})$ & $1.31^{+0.05}_{-0.06}$ & $1.23^{+0.05}_{-0.05}$ & $1.10^{+0.07}_{-0.06}$ & $1.02^{+0.07}_{-0.06}$ \\
$\Gamma$ & $2.03^{+0.02}_{-0.02}$ & $2.15^{+0.02}_{-0.02}$ & $2.04^{+0.02}_{-0.02}$ & $2.09^{+0.02}_{-0.02}$ \\
$\rm Log(\xi)$ & $3.60^{+0.09}_{-0.09}$ & $3.94^{+0.09}_{-0.10}$ & $3.70^{+0.10}_{-0.10}$ & $3.71^{+0.09}_{-0.10}$ \\
$\rm refl \_ frac$ & $0.25^{+0.05}_{-0.04}$ & $0.31^{+0.07}_{-0.06}$ & $0.31^{+0.07}_{-0.05}$ & $0.29^{+0.08}_{-0.05}$ \\
$\rm norm_{\rm r} \: (\rm \times 10^{-4})$ & $9.0^{+0.7}_{-0.6}$ & $13^{+2}_{-1}$ & $4.5^{+0.4}_{-0.4}$ & $6.8^{+0.7}_{-0.7}$ \\ \hline
$\chi^2 / \rm bins$ & $531.58/432$ & $400.80/417$ & $374.67/388$ & $398.39/401$ \\ \hline
$\chi^2 / \nu$ & \multicolumn{4}{c}{$1705.44/1602=1.06\:\:\:(1741^{+9}_{-8})$}
\enddata
\tablecomments{* represents a parameter fixed in the fit. In the row presenting the black hole spin, the number in parentheses represents the value that produced the lowest $\chi^2$ in the analysis. In the $\chi^2/\nu$ row, the number in parentheses represents the median $\chi^2$ value in the MCMC analysis with associated $\pm1\sigma$ variation.}
\end{deluxetable}

%%paragraph about the predictions of this model
This model produces similar constraints on the parameters as the best performing lamp post model.  The measured inclination of the accretion disk for model 18 is $Incl=33^{+7}_{-3}$ degrees, slightly larger than the one measured by model 6. The measured Fe abundance is consistent with the previous result, $A_{\rm Fe}=6^{+1}_{-1}$, despite the higher disk density. The temperature and normalization of the \texttt{diskbb} component are once again similar to the ones measured by model 6. The measurements of the power law index $\Gamma$ and ionization parameter of the accretion disk are once again slighltly lower, but show the same trend of increasing in the hard part of the spectra when compared to the soft part, in both observations. The reflection fraction parameter in the model also decreases to a value $\sim0.3$, once again consistent between the two observations. 

%%paragraph about emissivity
It was argued by \citealt{2012MNRAS.424.1284W} that the emissivity profiles in AGN have a steep power law index with radius between 6-8 for the inner regions of the accretion disk, which flattens for intermediate distances to almost a constant, and then tends to constant power law index of 3 over the outer regions of the disk. While current theoretical models do not yet have the capability of describing the emissivity of the illuminating source as a three component power law, the results of this model on the emissivity of the corona in an X-ray binary follow the same general trend: steep value $q_1=6\pm3$ for $r<12~r_g$ and a lower value $q_2=3\pm2$ for $r>12~r_g$. Models that link $q_1$ to $q_2$ for this data set perform poorly in terms of $\chi^2$ and have therefore not been explored further.

\section{Discussion} \label{sec:disc}

%% paragraphs summarizing results
We revisited the \textit{NuSTAR} observations of XTE J1908$+$094 during its 2013 and 2019 outburst. We divided each observation into two regimes, corresponding to an increase in hardness. Fitting the four data groups together using a set of 19 models that explore the entire set of physical assumptions that the \texttt{relxill} family of models covers returned consistent values of the black hole spin. The best performing model in terms of DIC (model 6) measures the black hole spin to have an intermediate value, $a = 0.55^{+0.29}_{-0.45}$. This model assumes a lamp post coronal geometry and fixes the height of the corona within the 2013 and 2019 observation. The corona appears to shift from $h_{\rm 2013}=14\pm2\;r_g$ in 2013 to $h=6\pm1\;r_g$ in 2019. The inclination of the inner accretion disk is low, $Incl=27^{+2}_{-3}$ degrees, the measured Fe abundance is $A_{\rm Fe}=6^{+2}_{-1}$ times the solar value, and the ionization of the accretion disk is high ($Log\;\xi\sim4$) during both observations, and increases with an increase in hardness. 

%% paragraph about flux, luminosity, eddington fraction
% For the 2013 observation of XTE J1908$+$094, the measured flux in the low and high hardness states was $F_{\rm 2013,low}=1.03\;(1.99)\;\times 10^{-9}\;erg/cm^2/s$ and $F_{\rm 2013,high}=1.19\;(2.17)\;\times 10^{-9}\;erg/cm^2/s$ in the $3-79\;(0.5-100)~keV$ band. For the 2019 observation, the measured fluxes are $F_{\rm 2019,low}=3.69\;(7.39)\;\times 10^{-10}\;erg/cm^2/s$ and $F_{\rm 2019,high}=4.52\;(8.33)\;\times 10^{-10}\;erg/cm^2/s$. 

For the 2013 observation of XTE J1908$+$094, the measured unabsorbed flux in the low and high hardness states was $F_{\rm 2013,low}=1.20\;(6.44)\;\times 10^{-9}\;erg/cm^2/s$ and $F_{\rm 2013,high}=1.36\;(6.77)\;\times 10^{-9}\;erg/cm^2/s$ in the 3--79~(0.5--100)~keV band. For the 2019 observation, the measured fluxes are $F_{\rm 2019,low}=0.42\;(2.78)\;\times 10^{-9}\;erg/cm^2/s$ and $F_{\rm 2019,high}=0.51\;(2.91)\;\times 10^{-9}\;erg/cm^2/s$. \citealt{2002A&A...394..553I} estimate the distance to XTE J1908$+$094 to be greater than 3~kpc based on its X-ray flux, while \citealt{2006MNRAS.365.1387C} constrain the distance to be between 3--10~kpc based on optical measurements. Using the flux measurement at the soft-to-hard X-ray state transition, \citealt{2015MNRAS.451.3975C} estimated the source distance to be between 4.8--8.3~kpc for a 3~$M_\odot$ black hole and between 7.8--13.6~kpc for an 8~$M_\odot$ black hole. Using these constraints on the distance to XTE J1908$+$098 and the mass of the central black hole, even during the highest flux phase (high hardness state of 2013 observation), the luminosity of the system is at most $\sim2.9\%$ of its Eddington luminosity when computed using the 3--79~keV flux or $\sim14.4\%$ when using the 0.5--100~keV flux. This maximal fraction is obtained for a $3\;M_\odot$ black hole at a distance of 8.3~kpc or a $8\;M_\odot$ black hole at 13.6~kpc. Even for the most conservative parameter combination which yields the maximum Eddington fraction, the value is well below the requirement of numerical simulations of $L/L_{Edd}\leq30\%$ for which the disk obeys the test particle ISCO (see, e.g., \citealt{2018ApJ...857....1F}). This indicates that the assumption of inner accretion disk extending all the way to the size of the ISCO radius is reasonable and that the inner disk regions are gravity dominated. During the faintest part of the observations (low hardness state of 2019 observation), the luminosity of the system is greater than $\sim0.3\%\;L_{Edd}$  using the 3--79~keV flux or $\sim1.9\%$ when using the 0.5--100~keV flux. These lower limits were obtained for a a $3\;M_\odot$ black hole at a distance of 4.8~kpc or a $8\;M_\odot$ black hole at 7.8~kpc. These lower limits are indicative of a soft-intermediate state, transitioning into a hard state.

% paragraph about the lessons learned: what changed, what improved, what needs to be done?
The 2013 data were previously analyzed by \citealt{2015ApJ...811...51T} and \citealt{2015ApJ...813...90Z}. Both papers indicated the presence of a relativistic broadened Fe K line, but the attempts to measure the black hole spin using relativistic reflection were unsuccessful. By fitting the 2013 and 2019 observations jointly, the effects of relativistic reflection become more discernible from the disk continuum radiation. Previous works rebinned the spectra requiring at least 50 counts per bin, while in this work, we used the optimal binning algorithm described by \citealt{2016A&A...587A.151K}. This binning scheme maintained the high signal/noise ratio at low energies, while allowing multiple bins at high energies. Therefore, for relatively low flux observations such as the ones analyzed in this paper, the optimal binning scheme allows the models to capture both the details of the strong, broadened Fe K line and accurately track the shape of the Compton hump at energies above $\sim20$~keV. The recent improvement in the \textit{NuSTAR} spectra extraction pipeline now accounts for the tear in the FPMA thermal blanket which used to lead to a flux difference between the two \textit{NuSTAR} sensors at low energies. All these factors in addition to the evolution of the \texttt{relxill} family of models in the recent times have encouraged pushing the potential of \textit{NuSTAR} observations to new frontiers and obtaining spin measurements from spectra that were previously considered unable to allow placing constraints on the black hole spin.

The \texttt{relxill} variants adopting the lamp post geometry produce consistent parameter constraints with the models that make no prior coronal geometry assumption, while maintaining a similar quality of the fit. For the spin measurement of EXO 1846-031, \citealt{2020ApJ...900...78D} found that \texttt{relxilllp} fails to measure the black hole spin due to the model requiring a large coronal height. The observation of EXO 1846-031 on which the measurement was made occurred when the source had a flux $\sim10$ times the flux of XTE J1908$+$094 during the 2013 observation and $\sim25$ times larger than during the 2019 observation. For lower signal/noise spectra, the lamp post models are able to produce similar quality fits as other model variants, and are preferred in terms of DIC due to their relative simplicity owed to a smaller number of free parameters. This analysis shows that despite the difference in assumption about the corona geometry, in relatively low signal/noise spectra the lamp post models are able to place similar spin constraints, despite not fully covering the same space of physical properties. For higher quality spectra such as the one presented in \citealt{2020ApJ...900...78D}, the data is additionally able to place constraints on the geometry of the corona and differentiate between lamp post and non-lamp post models.

% paragraph about increase in error bars from the old Miller et al. 2009
Using \textit{BeppoSAX} spectra, \citealt{2009ApJ...697..900M} estimated the spin of XTE J1908$+$094 to be $a = 0.75\pm0.09$. This analysis predicts the spin of XTE J1908$+$094 to be $a = 0.55^{+0.29}_{-0.45}$.  The two are formally consistent, but it is worth examining why the agreement is not better and why the errors are larger with better data. First, with the increase in energy resolution E/dE with \textit{NuSTAR} over \textit{BeppoSAX MECS} by a factor of nearly 2 at 6~keV and a factor of 4 at 10~keV, both the line profile and the shape of the continuum around the Fe K line can be more accurately constrained, reducing the possibility that the fit becomes stuck into a local minimum of $\chi^2$ in the parameter space as opposed to the global minimum, with falsely small errors. Second, in our fits to the \textit{NuSTAR} observations, we attempted to understand possible sources of systematic uncertainty by exploring the entire set of physical phenomena covered by the \texttt{relxill} family of models. We concluded that the choice of model does not have an important impact on the spin measurement, as the predictions of all models are consistent with each other.  However, the advances made in reflection models to include a wider range of physical effects has necessarily led to an increase in the number of free parameters in the models, leading to uncertainties that could not be probed in the past.  By including a broader set of physics, the \textit{accuracy} of the measurements has likely increased.  It should also be noted that the error scans in this work are more complete and conservative. By running an MCMC analysis on the entire set of parameters, the effects of all parameters are fully explored and incorporated into the uncertainty of the spin measurement. The larger and more physical uncertainty in the spin measurement has the potential to aid in eventually reconciling the spin distribution in X-ray binaries and BBH.
 
%% 1-2 paragraphs about spin distribution in XB and BBH. Compare, mention observational effects, posterior dependence on prior
The spin distribution of black holes in LMXB determined from relativistic reflection measurements has been estimated to have a mean around $a=0.66$ (\citealt{2015PhR...548....1M}). Based on the first two observing runs of Advanced LIGO and Advanced Virgo, it was estimated that half of the black holes in BBH systems have spins less than 0.27, and 90$\%$ have spins smaller than 0.55 (\citealt{2019ApJ...882L..24A}). While the posterior distribution of spin measurements from GW signals from BBH mergers is still strongly influenced by the assumed prior distribution, a better understanding of the spin distribution of black holes in X-ray binaries will significantly aid the effort of breaking the degeneracies between parameters. 

While works such as \citealt{2018A&A...614A..44K} have shown the robustness of supermassive black hole (SMBH) spin measurements in active galactic nuclei (AGN) using relativistic reflection, it is important to consider the effects of possible observational biases. One example is highlighted by \citealt{2016MNRAS.458.2012V} which argues that due to a selection effect, high spin SMBH are more likely to be detected and therefore measured. This effect is likely to also translate to black holes in X-ray binaries. Additionally, it was argued by \citealt{2021arXiv210403596J} that the mass range in the observed Galactic X-ray binaries is biased towards low mass black holes. The spins of black holes of different masses are likely to both be formed at different values and be influenced differently by evolution effects. For example, \citealt{2021arXiv210407493J} argued that small mass black holes might be subject to a stronger natal kick due to anisotropic mass ejection and neutrino emission, which can both increase the value of the birth black hole spin and alter the direction of the rotation axis. Additionally, using numerical GR MHD simulations of accretion, \citealt{2021arXiv210400741K} have shown that the accretion rate and black hole mass and spin evolution is strongly influenced by rotation dynamics of the infalling gas flow. Nevertheless, a black hole with an initial spin of $a = 0$ must accrete at least half its mass to reach a spin parameter of $a = 0.84$ (\citealt{1970Natur.226...64B}) and since stellar-mass black holes in LMXBs are more massive than their companion stars, it is unlikely that stellar-mass black holes can significantly alter their spin parameters on relatively short timescales.

\begin{figure}[ht]
    \centering
    \includegraphics[width=0.46\textwidth]{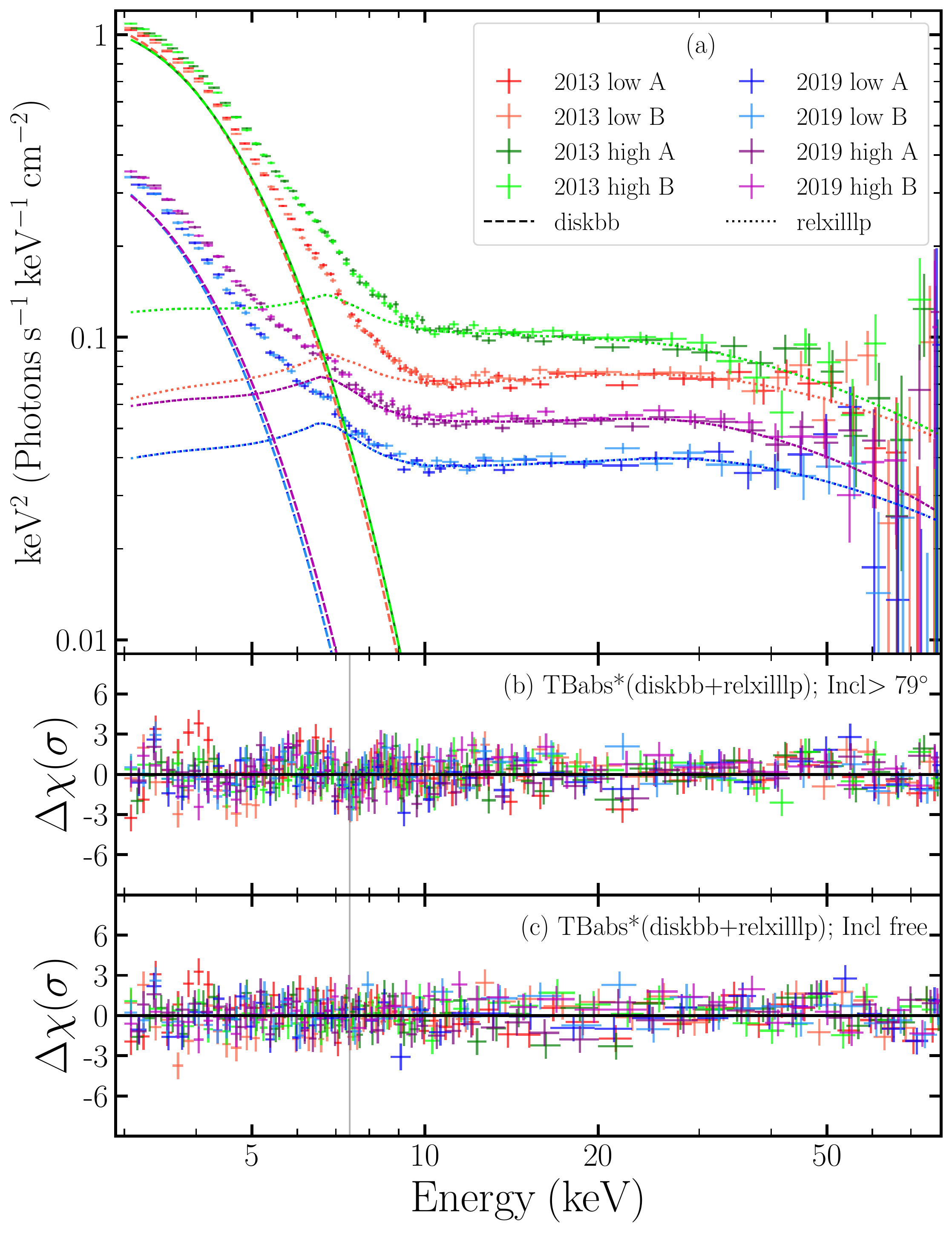}
    \caption{Panel (a) shows the unfolded spectrum of XTE J1908$+$094 when fit with model 6, requiring an inclination $>79^\circ$. The colors of the spectra match those described in Figure \ref{fig:delchi}. The dashed line shows the contribution of the \texttt{diskbb} component of the model, while the dotted line shows the contribution of the \texttt{relxilllp} component. Panels (b) and (c) show the residuals in terms of $\sigma$ of the fits using the high inclination and free inclination models, respectively. The $\chi^2$ of the model requiring a high inclination is $\sim85$ worse than the model that allows a free inclination. Note the increased residuals at $\sim7.4$~keV (indicated by the vertical gray line) in panel (b) when compared to panel (c), suggesting that the high inclination model overpredicts the shape of the blue wing of the Fe K line. Panel (c) is a copy of panel (d) in Figure \ref{fig:delchi}.}
    \label{fig:high_incl}
\end{figure}

The best performing model measures an inclination of the inner accretion disk of XTE J1908$+$094 of $27^{+2}_{-3}$ degrees, with all models predicting inclinations $<40^\circ$. In contrast, \citealt{2017MNRAS.468.2788R} measured an inclination larger than $79^\circ$ by analyzing the proper motion of expanding radio jets in this source under the assumption of two symmetric jets propagating ballistically outwards, with subsequent evolution dictated by light travel time delays and relativistic boosting.  Panel (a) in Figure \ref{fig:high_incl} shows the unfolded spectrum when fit with model 6, but requiring the inclination to be larger than $79^\circ$. The dashed lines represent the contributions of the \texttt{diskbb} component of the model, while the dotted line shows the contribution of the \texttt{relxilllp} component.  The $\chi^2$ becomes worse by $\sim85$ and the residuals (panel (b) of Figure \ref{fig:high_incl}) show that the large inclination model predicts an emission excess on the blue wing of the Fe K line. 

The residuals of the high inclination model shown in panel (b) in Figure \ref{fig:high_incl} could be interpreted as the absorption from a high velocity disk wind.  If we introduce a Gaussian component to our model with a negative amplitude and link its parameters within the 2013 and 2019 observation, the best fit $\chi^2$ improves, but is still worse than the case of the free inclination model by 21, for six additional parameters. In order for the $6.9$~keV absorption line of H-like Fe XXVI to be blue-shifted to the measured line center of the Gaussian absorption component at $7.4$~keV, the disk wind would have to travel at $v=0.07~c$. An outflow with that velocity would shift the $8.3$~keV line of He-like Fe XXV to $8.9$~keV. The absence of this feature in the residuals of the model would be indicative of very high ionization of the gas. 

Disk winds are typically observed in sources that are viewed at a high inclination.  Sharp flux dips are even more typical of sources viewed at high inclination.  The {\em lack} of dips in XTE J1908$+$094 and the requirement of a disk wind (of the same high velocity during the 2013 and 2019 outbursts) that perfectly balances the blue wing of the relativistic Fe K line -- represent a fortuitous combination of physical effects needed to explain a statistically disfavoured model. The low inclination model is then preferred both in terms of simplicity of the model and in terms of statistical significance, indicating that the large inclination determined by \citealt{2017MNRAS.468.2788R} was likely influenced by interactions of the radio jets with the ISM.  \citealt{2017MNRAS.468.2788R} note that the velocities implied by the jets do not match the flux differences between them, and that the techniques used to measure the inclination of jets in sources like GRS~1915$+$105 may not hold in XTE J1908$+$094.  In the future, simultaneous measurements of disk winds and relativistic reflection will be possible by combining \textit{NuSTAR} observations with \textit{XRISM}(\citealt{2018SPIE10699E..22T}).

\newpage

\acknowledgements
We thank James Miller-Jones and Elena Gallo for helpful discussions on the intriguing radio observations of XTE~J1908$+$094.  We thank Fiona Harrison, Karl Forster, and the NuSTAR team for executing observations of this source.  The authors also acknowledge Brandon Case and Linda Hudson for computer support that made this work possible during COVID restrictions. We also thank the anonymous referee for their helpful comments and suggestions, which have clarified and improved this paper. JAT acknowledges partial support under NASA NuSTAR Guest Investigator grant 80NSSC20K0644. ESK acknowledges financial support from the Centre National d’Etudes Spatiales (CNES).

\clearpage

\appendix

\section{Tested models} \label{sec:models}

Table \ref{table:models} shows the complete list of the 19 models tested in this analysis, together with the \texttt{relxill} variant they adopt, a brief summary of their assumptions, the best-fit $\chi^2$, number of degrees of freedom $\nu$, reduced $\chi^2$, computed DIC based on the MCMC runs, and the model rank in terms of lowest DIC.

\begin{deluxetable}{lllccccc}[ht]
\tablecaption{List of the 19 tested models \label{table:models}}
\tablewidth{\textwidth} 
\tabletypesize{\scriptsize}
\tablehead{
\colhead{Model} & \colhead{\texttt{relxill} component} & \colhead{Assumptions} & \colhead{$\chi^2$} & \colhead{$\nu$} & 
 \colhead{$\chi^2/\nu$} & \colhead{DIC} & \colhead{Rank (by DIC)} 
}
\startdata\\
\multicolumn{1}{l|}{1} & relxilllp & linked diskbb norm, free h & 1726.39 & 1605 & 1.076 & 1810.25 & 13 \\
\multicolumn{1}{l|}{2} & relxilllp & linked diskbb norm, linked h within observation & 1731.16 & 1607 & 1.077 & 1787.72 & 8 \\
\multicolumn{1}{l|}{3} & relxilllp & linked diskbb norm, linked all h & 1779.16 & 1608 & 1.106 & 1846.02 & 17 \\
\multicolumn{1}{l|}{4} & relxilllp & linked diskbb norm, T, and h within observation & 1728.36 & 1608 & 1.075 & 1797.4 & 11 \\
\multicolumn{1}{l|}{5} & relxilllp & free diskbb norm, free h & 1707.56 & 1602 & 1.066 & 1784.94 & 6 \\
\multicolumn{1}{l|}{6*} & relxilllp & free diskbb norm, linked h within observation & 1707.44 & 1604 & 1.064 & 1762.7 & 1 \\
\multicolumn{1}{l|}{7} & relxilllp & free diskbb norm, linked all h & 1720.16 & 1605 & 1.072 & 1774.06 & 2 \\
\multicolumn{1}{l|}{8} & relxill & free diskbb norm, free q1,q1,r$\_$br & 1709.62 & 1594 & 1.073 & 1811.62 & 14 \\
\multicolumn{1}{l|}{9} & relxill & r$\_$br fixed at 12 & 1706.26 & 1598 & 1.068 & 1786.95 & 7 \\
\multicolumn{1}{l|}{10} & relxill & linked diskbb norm, free q1,q2,r$\_$br & 1714.83 & 1597 & 1.074 & 1821.33 & 16 \\
\multicolumn{1}{l|}{11} & relxillD & N=15 free diskbb norm, free q1,q2,r$\_$br & 1708.45 & 1598 & 1.069 & 1800.79 & 12 \\
\multicolumn{1}{l|}{12} & relxillD & N=15, r$\_$br=12 & 1710.35 & 1602 & 1.068 & 1781.03 & 4 \\
\multicolumn{1}{l|}{13} & relxillD & N=15 linked diskbb norm, free q1,q2,r$\_$br & 1724.13 & 1601 & 1.077 & 1866.92 & 18 \\
\multicolumn{1}{l|}{14} & relxillD & N=17 free diskbb norm, free q1,q2,r$\_$br & 1708.59 & 1598 & 1.069 & 1818.26 & 15 \\
\multicolumn{1}{l|}{15} & relxillD & N=17, r$\_$br=12 & 1710.53 & 1602 & 1.068 & 1783.98 & 5 \\
\multicolumn{1}{l|}{16} & relxillD & N=17 linked diskbb norm, free q1,q2,r$\_$br & 1723.25 & 1601 & 1.076 & 1880.98 & 19 \\
\multicolumn{1}{l|}{17} & relxillD & N=19 free diskbb norm, free q1,q2,r$\_$br & 1705.02 & 1598 & 1.067 & 1789.01 & 9 \\
\multicolumn{1}{l|}{18*} & relxillD & N=19, r$\_$br=12 & 1705.44 & 1602 & 1.065 & 1778.81 & 3 \\
\multicolumn{1}{l|}{19} & relxillD & N=19 linked diskbb norm, free q1,q2,r$\_$br & 1711.81 & 1601 & 1.069 & 1789.51 & 10
\enddata
\tablecomments{Models 6 and 18 (marked by *) are the best performing models in terms of DIC among those adopting the lamp post geometry and those making no prior coronal geometry assumptions, respectively. The two models are discussed in detail in this work. Model 18 performs best out of the models using \texttt{relxill} or \texttt{relxilllp}, but ranks only third among all tested models.}
\end{deluxetable}

\clearpage

\section{Corner plot for model 6} \label{sec:corner}

\begin{figure}[ht]
    \centering
    \includegraphics[width=0.9\textwidth]{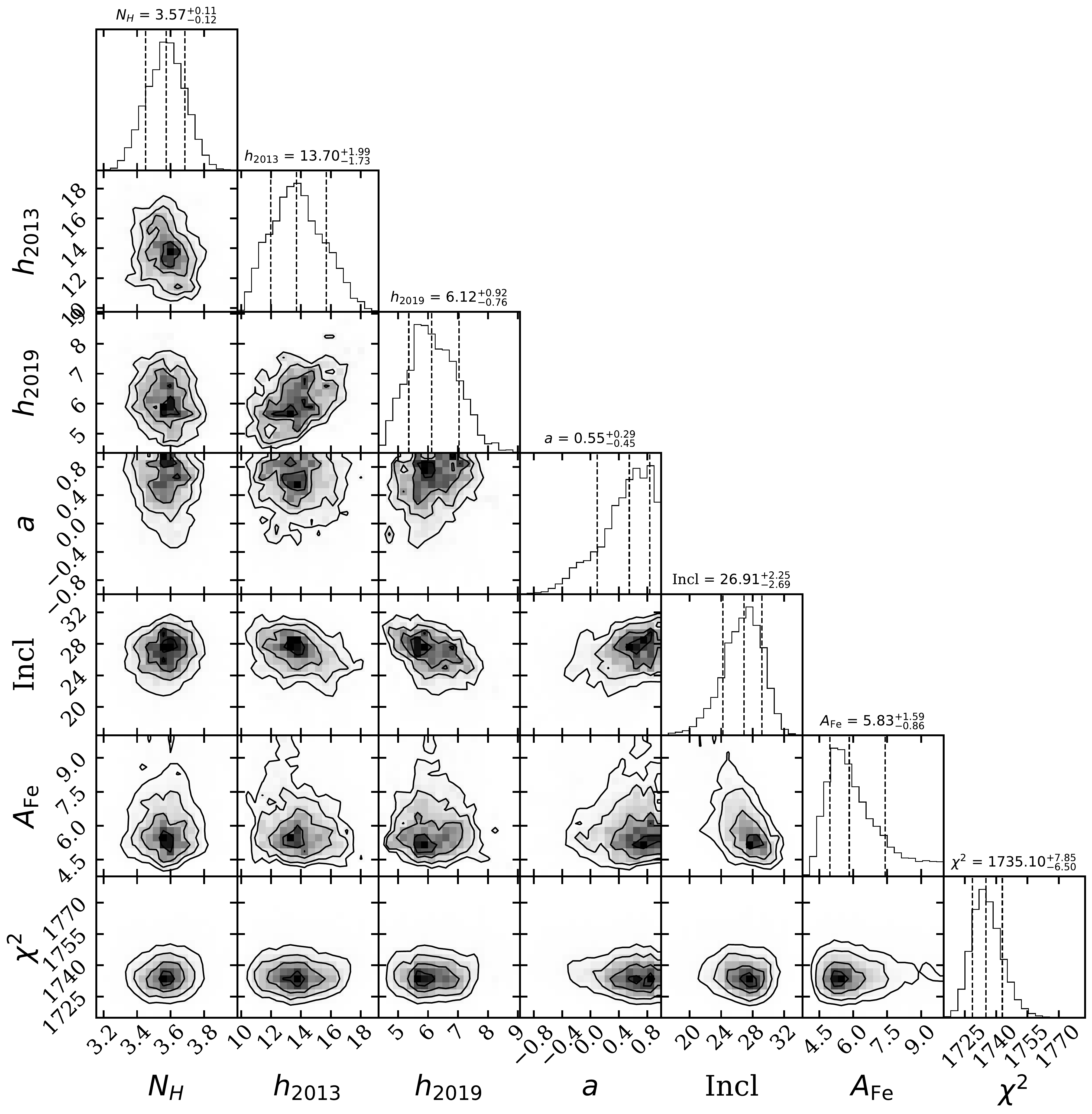}
    \caption{Corner plot of some parameters of interest in model 6. The parameters shown are the interstellar hydrogen column $N_H$ in units of $10^{22}\;cm^{-2}$, the height of the corona in the lamp post geometry in the 2013 and 2019 observations in units of $r_g$, the black hole spin $a$, inclination of the inner accretion disk in degrees, the Fe abundance $A_{\rm Fe}$ in solar units, and the $\chi^2$ distribution.}
    \label{fig:corner_small}
\end{figure}

Figure \ref{fig:corner_small} shows the corner plot of a few parameters of interest in model 6. The parameters presented are the equivalent hydrogen column along the line of sight $N_H$ in units of $10^{22}\;cm^{-2}$, the height of the corona in the lamp post geometry in the 2013 and 2019 observations in units of $r_g$, the black hole spin $a$, inclination of the inner accretion disk in degrees, the Fe abundance $A_{\rm Fe}$ in solar units, and the $\chi^2$ distribution. 

\clearpage
\bibliography{paper}{}
\bibliographystyle{aasjournal}

\end{document}